\documentstyle[prl,aps,epsfig]{revtex}
\input rotate
\newcommand{\bfm}[1]{\mbox{\boldmath${#1}$}}
\begin{document}
\draft
\title{Quantum vortices in systems obeying a \\generalized exclusion principle}
\author{G. Kaniadakis \footnote{Electronic address:kaniadakis@polito.it}, A. M. Scarfone \footnote{Electronic address: scarfone@polito.it}}

\address{\footnotesize Dipartimento di Fisica, Politecnico di Torino,
Corso Duca degli Abruzzi 24, I-10129 Torino, Italy\\ and Istituto
Nazionale per la Fisica della Materia, Unit\'a del Politecnico di
Torino}
\date{\today}
\maketitle
\begin {abstract}
The paper deals with a planar particle system obeying a
generalized exclusion principle (EP) and governed, in the mean
field approximation, by a nonlinear Schr\"odinger equation. We
show that the EP involves a mathematically simple and physically
transparent mechanism, which allows the genesis of quantum
vortices in the system. We obtain in a closed form the shape of
the vortices and investigate its main physical properties.

\end {abstract}
\vspace{6mm}

In literature experimental evidences has been presented of
formation of quantized vortex structures in different macroscopic
quantum systems. Vortices have been observed both in Bose
(Bose-Einstein condensate of alkali atom clouds) and in Fermi
($^3$He-A superfluids, heavy fermion superconductors UPt$_3$ and
U$_{0.97}$Th$_{0.03}$Be$_{13}$) systems
\cite{Williams,Matthews,Madison,Heffuer,Zieve,Shung}. Actually,
there exist different theoretical models admitting stationary
solutions of vortex type
\cite{Lund,Rica,Koplik,Ivonin,Josse,Jackiw,Bara,Abra,Has,Papa,Str}.
It is apparent that the vortex type solution is imposed by the non
linearities introduced in the particular model adopted. For
instance, by considering a non relativistic matter system, one can
construct different models starting from Non Linear Schr\"odinger
Equations (NLSEs) or from the gauged NLSEs in the frame of a
Maxwell or Chern-Simons theory.

Principal goal of the present work is to show that there exists a
mathematically simple and physically transparent mechanism,
imposed by the generalized Exclusion Principle (EP), which allows
the genesis of quantum vortices in planar non relativistic
particle systems.

Since 1932, it was clear that the effects due to the statistics
and imposed to a system of free fermions by the Pauli exclusion
principle  can be simulated by a repulsive potential in the
coordinate space \cite{Uhlenbeck}. In the same way for bosons, an
attractive potential can simulate the statistical behaviour of the
system. After 1940, in different works, particle systems obeying
to statistics, which are different from the standard Bose-Einstein
and Fermi-Dirac ones, have been considered
\cite{Gentile,Green,Greenberg,Haldane}.

Recently, it has been studied a many body quantum system obeying a
generalized Exclusion-Inclusion Principle (EIP)
\cite{KQ,Kadanoff,Frank,KPRA,KQSPRE,KQSRMP}. The peculiar property
of this system is that the EIP introduces an attractive or
repulsive potential in the coordinate space and can simulate the
intermediate (between bosonic and fermionic) behaviour of the
system. We consider some examples of real physical systems where
EIP can be usefully applied. The Bose-Einstein condensation
originates from an attraction of statistical nature (Bose-Einstein
statistics) among the particles. In several systems the
Bose-Einstein condensation is studied by means of a cubic NLSE
which describes in mean field approximation an attractive
interaction between two bodies. In place of the cubic and simplest
interaction, other interactions can be considered as, for
instance, the one introduced by EIP to simulate an attraction
among the particles. Analogously, in the case of superfluids or
semiconductors Fermionic systems, the repulsive interaction among
the particles originated from the Fermi-Dirac statistics can be
simulated by the EIP with $\kappa<0$. In condensed matter the
motion of the couple electron-hole in a semiconductor can be
described again by EIP. In fact, while electrons and holes are
fermions, together can be considered excited states behaving
differently from a fermion or a boson. Moreover, in nuclear
physics the interaction among the fermionic valence nucleons
outside the core produces pairs of correlated nucleons that can be
approximated as particles with a behaviour intermediate between
fermionic and bosonic ones. This nuclear state (quasideuteron
state) can be viewed as a particle system that obeys to EIP.\\ The
dynamics of the canonical quantum system obeying the EIP is
governed by the following NLSE:
\begin{eqnarray}
i\hbar\frac{\partial\psi}{\partial
t}=-\frac{\hbar^2}{2\,m}\Delta\psi+W(\rho,{\bfm j})\psi+i\,{\cal
W}(\rho,{\bfm j})\psi+V\psi \ ,\label{NSE}
\end{eqnarray}
where the real and imaginary parts of the nonlinearity are given
respectively by:
\begin{eqnarray}
&&W(\rho,\,{\bfm j})=\kappa\,\frac{m}{\rho}\left(\frac{\bfm
j}{1+\kappa\,\rho}\right)^2 \ ,\\ &&{\cal W}(\rho,\,{\bfm
j})=-\kappa\,\frac{\hbar}{2\,\rho}\,{\bfm\nabla}\!\cdot\!\left(\frac{{\bfm
j}\,\rho}{1+\kappa\,\rho}\right) \ .
\end{eqnarray}
The free parameter $\kappa$ is a constant which takes into account
the intensity of the exclusion-inclusion statistical effects. It
is easy to verify that the system described by Eq. (\ref{NSE})
obeys the  continuity equation
\begin{eqnarray}
\frac{\partial\rho}{\partial t}+{\bfm\nabla}\cdot{\bfm j}=0 \
,\label{ce}
\end{eqnarray}
where $\rho=|\psi|^2$, while the quantum current $\bfm j$ is given
by:
\begin{eqnarray}
{\bfm
j}=-\frac{i\,\hbar}{2\,m}\,(1+\kappa\,\rho)\,(\psi^\ast\,{\bfm\nabla}\psi
-\psi\,{\bfm\nabla} \psi^\ast) \ .\label{current}
\end{eqnarray}
Eq. (\ref{ce}) assures the conservation of the particle number
$N=\int\rho\,d^Dx$ of the system.

We discuss now briefly the origin of the model described by Eq.
(\ref{NSE}). We start by considering the following nonlinear
Fokker-Planck equation \cite{KQ,Kadanoff}:
\begin{eqnarray}
\frac{\partial\,\rho}{\partial\,t}+{\bfm\nabla}\cdot\left[\frac{{\bfm
\nabla}\,S}{m}\,\rho\,(1+\kappa\,\rho)+D\,{\bfm\nabla}\,\rho\right]=0
\ ,\label{FP}
\end{eqnarray}
being ${\bfm\nabla}\,S/m$ the drift velocity and $D$ the diffusion
coefficient. In the case $\kappa=0$ and $D=0$, the above
Fokker-Planck equation reduces to the well-known continuity
equation for $\rho=|\psi|^2$ and the linear quantum mechanics can
be obtained if we use the ansatz
\hbox{$\psi=\rho^{1/2}\,\exp(i\,S/\hbar)$}. In Ref. \cite{Doebner}
has been considered the case $\kappa=0,\,\,D\not=0$ and starting
from Eq. (\ref{FP}) a new NLSE was obtained. Differently, starting
from Eq. (\ref{FP}), after posing $\kappa\not=0$ and $D=0$, Eq.
(\ref{NSE}) can be obtained.

The introduction in Eq. (\ref{NSE}) of the factor $1+\kappa\,\rho$
originates from the presence of the EIP and allows us to take into
account many particle quantum effects. In fact, the transition
probability from the site $\bfm x$ to $\bfm x^\prime$ is defined
as $\pi(t,\,\bfm x\rightarrow\bfm x^\prime)=r(t,\,\bfm x,\,\bfm
x^\prime)\,\rho(t,\,\bfm x)\,[1+\kappa\,\rho(t,\,\bfm x^\prime)]$
with $r(t,\,\bfm x,\,\bfm x^\prime)$ the transition rate. The
transition probability depends on the particle population
$\rho(t,\,\bfm x)$ of the starting point $\bfm x$, and also on the
population $\rho(t,\,\bfm x^\prime)$ of the arrival point $\bfm
x^\prime$. For $\kappa\not=0$ the EIP holds and the parameter
$\kappa$ quantifies how much the particle kinetics is affected by
the particle population of the arrival point. If $\kappa>0$ the
$\pi(t,\,\bfm x\rightarrow\bfm x^\prime)$ introduces an inclusion
principle. In fact the population at the arrival point $\bfm
x^\prime$ stimulates the transition, and the transition
probability increases linearly with $\rho(t,\,\bfm x^\prime)$. In
the case $\kappa<0$ the $\pi(t,\,\bfm x\rightarrow\bfm x^\prime)$
takes into account the Pauli exclusion principle. If the arrival
point $\bfm x^\prime$ is empty, $\rho(t,\,\bfm x^\prime)=0$, and
the $\pi(t,\,\bfm x\rightarrow\bfm x^\prime)$ depends only on the
population of the starting point. If the arrival site is populated
$0<\rho(t,\,\bfm x^\prime)\leq\rho_{\rm max}$, and the transition
is inhibited. The range of values the parameter $\kappa$ can
assume is bounded by the condition that $\pi(t,\,\bfm
x\rightarrow\bfm x^\prime)$ be real and positive, as $r(t,\,\bfm
x,\,\bfm x^\prime)$ is. Thus we conclude that
$\kappa\geq-1/\rho_{\rm max}$.

The form of the nonlinearity ${\cal W}(\rho,{\bfm j})$ in
(\ref{NSE}) is imposed by the continuity equation (\ref{ce}),
while the form of $W(\rho,{\bfm j})$ is imposed by the requirement
of the canonicity of the system. The Hamiltonian density of the
system (\ref{NSE}) is given by:
\begin{eqnarray}
{\cal
H}=\frac{\hbar^2}{2\,m}\,\left|{\bfm\nabla}\psi\right|^2+U_{_{\rm
EIP}}+V\,\rho \ , \label{HD}
\end{eqnarray}
being
\begin{eqnarray}
U_{_{\rm EIP}}=\kappa \,\rho ^2 \, \frac{({\bfm\nabla}S)^2}{2\,m}
\ ,
\end{eqnarray}
the nonlinear potential introduced by the EIP. In Ref.
\cite{KQSPRE} it has been shown that the system (\ref{NSE}) admits
1-D solitons.

Following the standard procedure of the linear quantum mechanics,
we define the quantum velocity ${\bfm v}$ through ${\bfm
j}\!=\!{\bfm v}\rho$. Taking into account Eq. (\ref{current}) and
writing $\psi$ in terms of the hydrodynamic variables
\hbox{$\psi=\rho^{1/2}\,\exp(i\,S/\hbar)$}, we have:
\begin{eqnarray}
{\bfm v}=(1+\kappa\,\rho)\,\frac{{\bfm\nabla}S}{m} \ .
\end{eqnarray}
The quantum velocity can be expressed in terms of the Clebsch
potentials $S,\lambda,\mu$ through $m\bfm{v}=\bfm{\nabla}S+
\lambda {\bfm\nabla}\mu$. It results that the EIP imposes the
choice $\lambda=\kappa \rho$ and $\mu=S$ for the Clebsch
potentials. Now, we define the vorticity $\bfm \omega$ of the
system, through ${\bfm \omega}={\bfm\nabla}\times{\bfm v}$ and
obtain
\begin{eqnarray}
{\bfm \omega}=\frac{\kappa}{m}\, {\bfm\nabla}\rho \times {\bfm
\nabla}S \ . \label{vor}
\end{eqnarray}
From (\ref{vor}), it results that Eq. (\ref{NSE}) describes a
vorticose system. In the present contribution, we will consider
the planar, $D=2$, static vortex solutions of Eq. (\ref{NSE}) in
the case $\kappa=-\xi<0$, when EIP is reduced to an exclusion
principle (EP). By introducing the plane polar coordinates
$r=\sqrt{x^2+y^2}$ and $\theta=\arctan(y/x)$ in the plane, we
search for solutions of Eq. (\ref{NSE}) in which $S=S(\theta)$ and
$\rho=\rho(r)$. The continuity equation imposes $S=\hbar\,n\,
\theta$ and consequently we write $\psi$ as:
\begin{eqnarray}
\psi(r,\,\theta)=\rho(r)^{1/2}\,e^{i\,n \,\theta} \
.\label{ansatz}
\end{eqnarray}
The parameter $n$ must be integer in order to make $\psi$, given
by Eq. (\ref{ansatz}), a single value function. In the following,
we impose $\int\rho\,d^2x=N$, and therefore $\rho (\infty)=0$. Let
us note that Eq. (\ref{NSE}) is not Galilei invariant
\cite{KQSRMP} so that traveling solutions can not be obtained by
boosting static solutions of Eq.(\ref{NSE}).

The quantum velocity $\bfm v$ for the vortices (\ref{ansatz})
becomes:
\begin{eqnarray}
{\bfm v}=\frac{\hbar}{m}\,\frac{n}{r}\,(1-\xi\,\rho)\, {\hat{\bfm
e}} _{_{\theta}} \ , \label{vq}
\end{eqnarray}
where $\hat{\bfm e}_{_{\theta}}$ is the unitary vector orthogonal
to the vector ${\bfm r}=(x,\,y)$.  After integration of Eq.
(\ref{vq}) on the circle $\gamma_{_\infty}$ with center at the
vortex core and with a radius $R\rightarrow\infty$, we obtain the
following relevant property:
\begin{eqnarray}
m\,\oint_{\gamma_{_\infty}}{\bfm v}\cdot
d{\bfm\ell}=2\,\pi\,\hbar\,n \ , \label{G8}
\end{eqnarray}
which justifies the name of vorticity index $n$. The vorticity
$\bfm{\omega}=\omega\,\hat{\bfm e}_{_{\!z}}$ of the system is
given by:
\begin{eqnarray}
\omega=\frac{2\,\pi\,\hbar\,n}{m}\,\delta^2({\bfm
r})-\xi\frac{\hbar}{m}\,\frac{n}{r}\,\frac{ d \rho}{d r} \ ,
\label{vort}
\end{eqnarray}
and taking into account $\rho(\infty)=0$, it is easy to verify
that the total vorticity depends exclusively on the behaviour of
the vortex core
\begin{eqnarray}
\int {\omega}\,d^2x=2\,\pi\,n\,\frac{\hbar}{m} \ . \label{TV}
\end{eqnarray}
The definition of the vorticity ${\bfm
\omega}={\bfm\nabla}\times{\bfm v}$, imposes the relation:
\begin{eqnarray}
\int {\cal \omega}\,d^2x=\oint_{\gamma_{_\infty}}{\bfm v}\cdot
d{\bfm\ell} \ ,
\end{eqnarray}
which can be easily verified by comparing Eqs (\ref{G8}) and
(\ref{TV}).

The angular momentum of the vortex, which is a vector orthogonal
to the vortex plane, can be calculated as mean value of the
operator $\hat{L}_z=-i\,\hbar\,\partial/\partial\theta$ and
assumes the following quantized value:
\begin{eqnarray}
L_z=n\,N\,\hbar \  \ , \label{angular}
\end{eqnarray}
which does not depend on the parameter $\xi$.

We can calculate now the complex non linearity $W+i\,{\cal W }$ in
Eq. (\ref{NSE}) for the vortex given by Eq. (\ref{ansatz}). The
non linearity becomes a real one and Eq. (\ref{NSE}) reduces to:
\begin{eqnarray}
i\,\hbar\,\frac{\partial\psi}{\partial
t}=-\frac{\hbar^2}{2\,m}\,\Delta\psi
-\frac{\hbar^2}{m}\,\,\frac{\xi\,n^2}{r^2}\, \rho \,\,\psi + V
\,\,\psi \ . \label{FQNSE}
\end{eqnarray}
Eq. (\ref{FQNSE}) contains a nonlinearity very close to the one of
cubic NLSE and describes a canonical quantum system with a
Hamiltonian density given by Eq. (\ref{HD}), where the nonlinear
potential introduced by the EP assumes the form:
\begin{eqnarray}
U_{_{\rm
EP}}=-\frac{\hbar^2}{2\,m}\,\,\frac{\xi\,n^2}{r^2}\,\rho^2 \ .
\label{U}
\end{eqnarray}

As regard the spatial shape of the vortex $\rho=\rho(r)$, we
insert Eq. (\ref{ansatz}) into Eq. (\ref{FQNSE}), and obtain the
following second order ordinary differential equation:
\begin{eqnarray}
&&\frac{1}{r\,\rho}\,
\frac{d}{dr}\left(r\,\frac{d\rho}{dr}\right)-\frac{1}{2}\left(\frac{1}{\rho}\,
\frac{d\rho}{dr}\right)^2 \nonumber \\
&&-\frac{2\,n^2}{r^2}\,(1-2\,\xi\,\rho)-\frac{4\,m}{\hbar^2}
V(r)=0 \ .\label{uno}
\end{eqnarray}
In the following, we are interested to study the free vortices
$V=0$. After introducing the dimensionless variable
\begin{eqnarray}
z=2\,n\,\log\,\frac{r}{r_n} \ ,
\end{eqnarray}
where $r_n$ is an arbitrary constant,  Eq. (\ref{uno}) becomes:
\begin{eqnarray}
{2\over\rho}\,\frac{d^2\rho}{d z^2}- \,\left({1\over\rho}
\,\frac{d\rho}{dz}\right)^2+ 2\,\xi\,\rho-1=0 \ .\label{due}
\end{eqnarray}
Now we consider the auxiliary function $y(\rho)$
\begin{eqnarray}
y(\rho)=\left({1\over\rho}\,\frac{d\rho}{dz}\right)^2 \
,\label{trasf}
\end{eqnarray}
which, taking into account Eq. (\ref{due}), obeys the differential
equation
\begin{eqnarray}
\frac{dy}{d\rho}+\frac{y}{\rho}-\frac{1}{\rho}\,(1-2\,\xi\,\rho)=0
\ . \label{DUE1}
\end{eqnarray}
The function $y(\rho)$ after integration of (\ref{DUE1}) assumes
the form
\begin{eqnarray}
y(\rho)=\frac{\alpha}{\rho}+ 1-\xi\,\rho \ ,\label{tre}
\end{eqnarray}
being $\alpha$ an integration constant. Combining Eq.
(\ref{trasf}) and Eq. (\ref{tre}), we arrive to the following
first order ordinary differential equation for the shape of the
vortex:
\begin{eqnarray}
\left(\frac{d\rho}{dz}\right)^2=\alpha \,\rho+
\rho^2\,(1-\xi\,\rho) \ . \label{quattro}
\end{eqnarray}
\begin{figure}

{{\hspace*{-1.7cm}\rotstart{0 rotate
}{\scalebox{.45}{\includegraphics{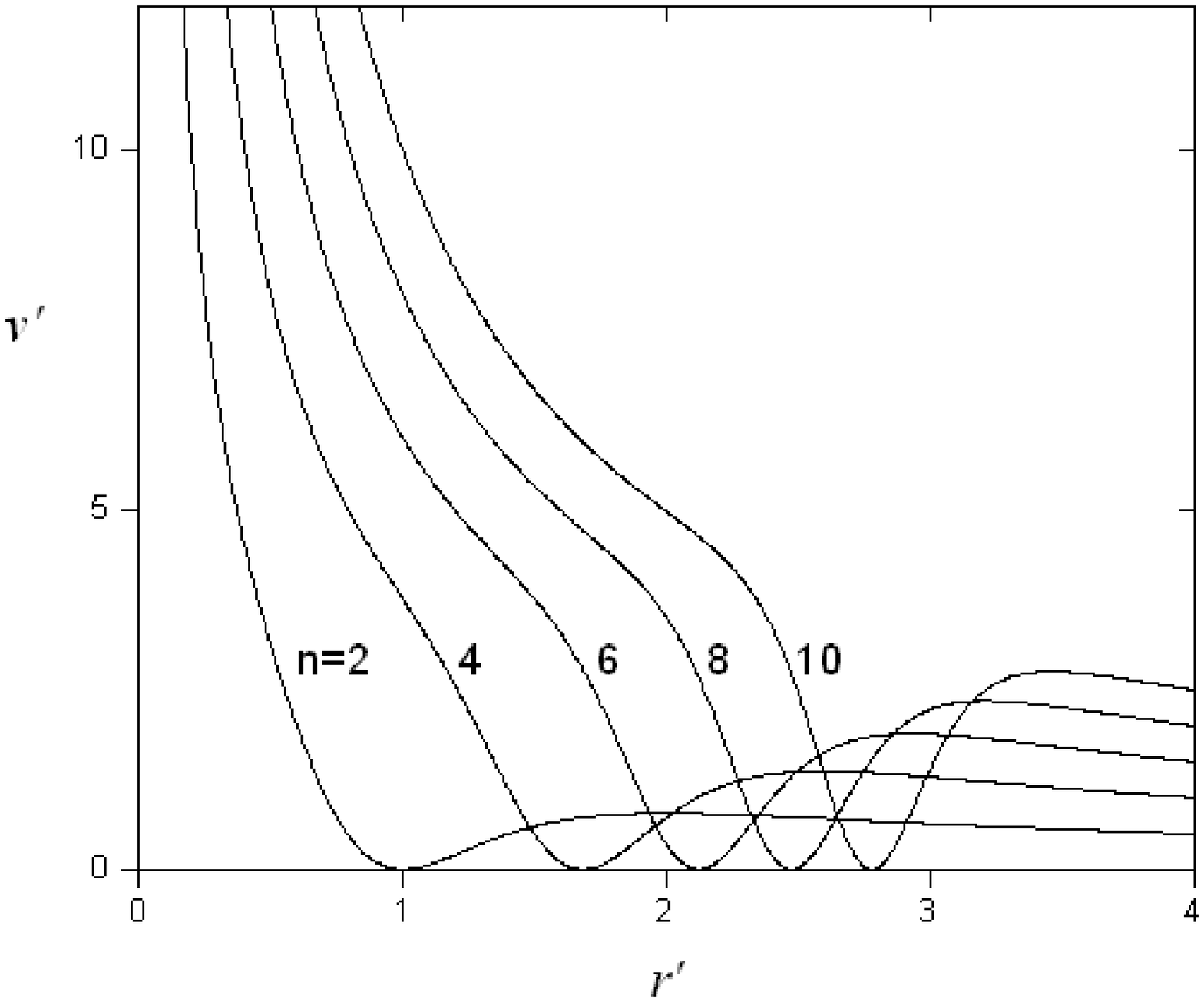} }}\rotfinish}}
{\vspace*{-8.7cm}} {{\hspace*{8cm}\rotstart{0 rotate
}{\scalebox{.45}{\includegraphics{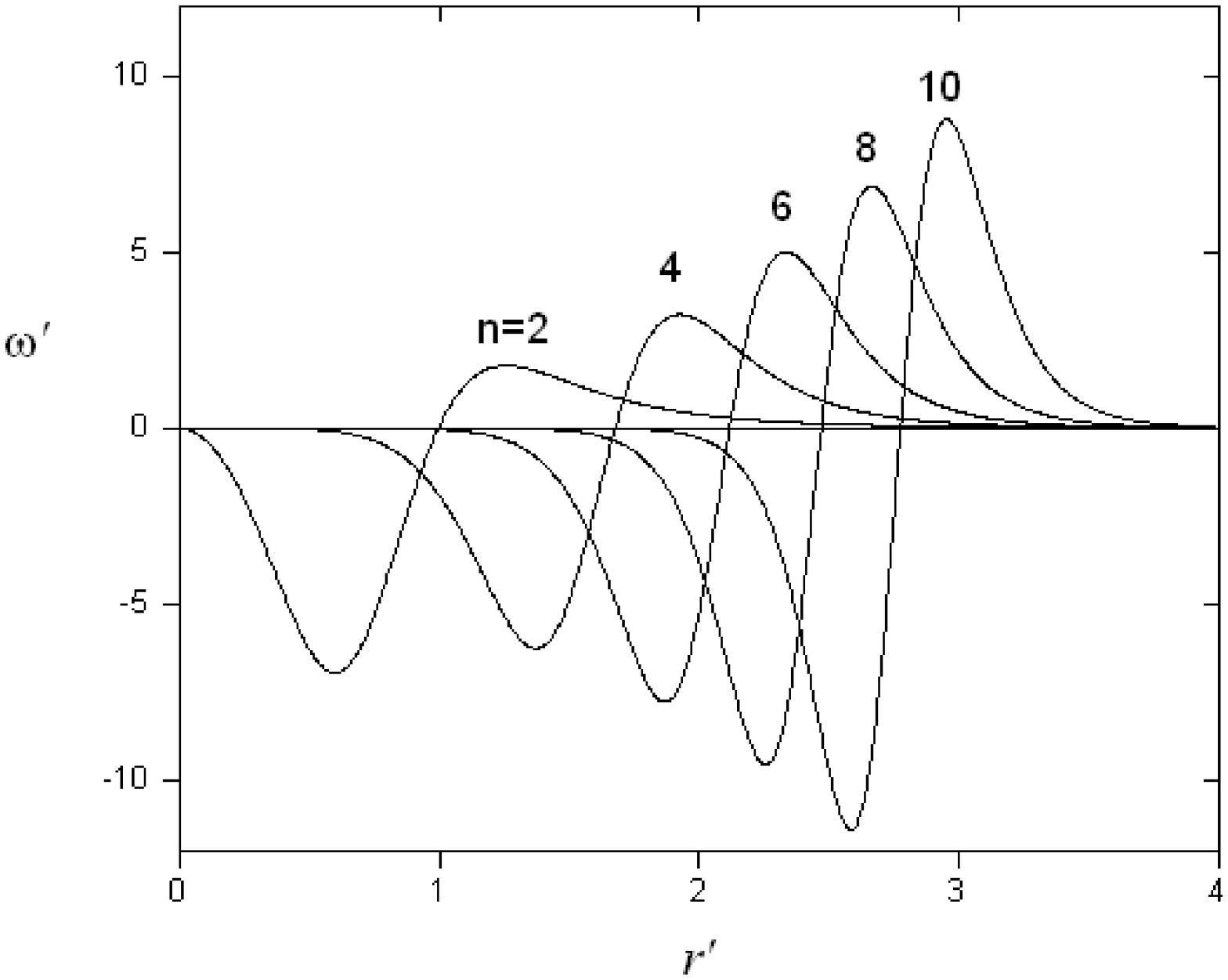} }}\rotfinish}}


\begin{minipage}[t]{8cm}\caption{{\footnotesize Plot of the dimensionless quantum
velocity $v^\prime=v\,m\,r_2/\hbar$ with $r_2=\sqrt{\xi\,N}/\pi$
as a function of the dimensionless distance from the vortex core
$r^\prime=r/r_2$ for the vortices with $n=2,4,6,8,10$.
}}\end{minipage} \hspace{1cm}
\begin{minipage}[t]{8cm}\caption{{\footnotesize Plot of the non singular part of the dimensionless
vorticity $\omega^\prime=\omega\,m\,r_2^2/\hbar$ with
$r_2=\sqrt{\xi\,N}/\pi$ as a function of the dimensionless
distance from the vortex core $r^\prime=r/r_2$ for the vortices
with $n=2,4,6,8,10$. }}
\end{minipage}
\end{figure}
All the physical solutions of Eq. (\ref{quattro}), whatever the
value of the parameter $\alpha\in I\!\!R$, must be non singular,
non negative, and normalizable.

The Hamiltonian $H=\int {\cal H}\,d^2x$ of the system can be
calculated starting from (\ref{HD}) and the expression (\ref{U})
for $U_{_{\rm EP}}$:
\begin{eqnarray}
H=\frac{\hbar^2n^2}{m}\int \left[\frac{\alpha}{2}\,{1\over
r^2}+\frac{1}{r^2}\,\rho\,(1-\xi\,\rho)\right]\,d^2x \
.\label{ham}
\end{eqnarray}
We must choose $\alpha=0$ in order to have a finite value of the
Hamiltonian. After integration on the variable $\theta$, we obtain
the following simple expression for the Hamiltonian
\begin{eqnarray}
H=2\,\pi\,n\,\hbar\int^\infty_0 j\,dr \ , \label{HJ}
\end{eqnarray}
being $j=(\hbar n/m\,r)\rho (1-\xi \rho)$ the value of the current
\hbox{${\bfm j}=j\, {\hat{\bfm e}} _{_{\theta}}$} given by Eq.
(\ref{current}). Finally, also the differential equation for the
vortex shape assumes a very simple form
\begin{eqnarray}
\left(\frac{d\rho}{dz}\right)^2=\rho^2\,(1-\xi\,\rho) \ ,
\label{EDV}
\end{eqnarray}
and can be easily integrated, obtaining the vortex profile in an
explicit form
\begin{eqnarray}
\rho(r)=\frac{4}{\xi}\,\left[\left(\frac{r}{r_n}\right)^n+
\left(\frac{r_n}{r}\right)^n\right]^{-2} \ .\label{sol1}
\end{eqnarray}
Therefore, the wave function of the vortex becomes
\begin{eqnarray}
\psi(r,\,\theta)=\frac{2}{\sqrt{\xi}}\,\left[\left(\frac{r}{r_n}
\right)^n+ \left(\frac{r_n}{r}\right)^n\right]^{-1}\exp\,
({i\,n\,\theta}) \ . \label{psivortex}
\end{eqnarray}
The free parameter $r_n$, which absorbs the integration constant
related to Eq. (\ref{EDV}), can be calculated from the
normalization condition  $2\,\pi\,\int^\infty_0\rho(r)\,r\,dr=N$
and assumes the value
\begin{eqnarray}
\omega=\frac{2\,\pi\,\hbar\,n}{m}\,\delta^2({\bfm
r})-\frac{8\,n^2\,\hbar}{m\,r_n^2}\,\left(\frac{r}{r_n}\right)^{2\,(n-1)}
\,\left[1-\left(\frac{r}{r_n}\right)^{2\,n}\right]\,\left[1+\left(\frac{r}{r_n}\right)
^{2\,n}\right]^{-3} \ .\label{shape1}
\end{eqnarray}
\begin{figure}
\vspace*{-1cm}

{{\hspace*{-1.7cm}\rotstart{0 rotate
}{\scalebox{.45}{\includegraphics{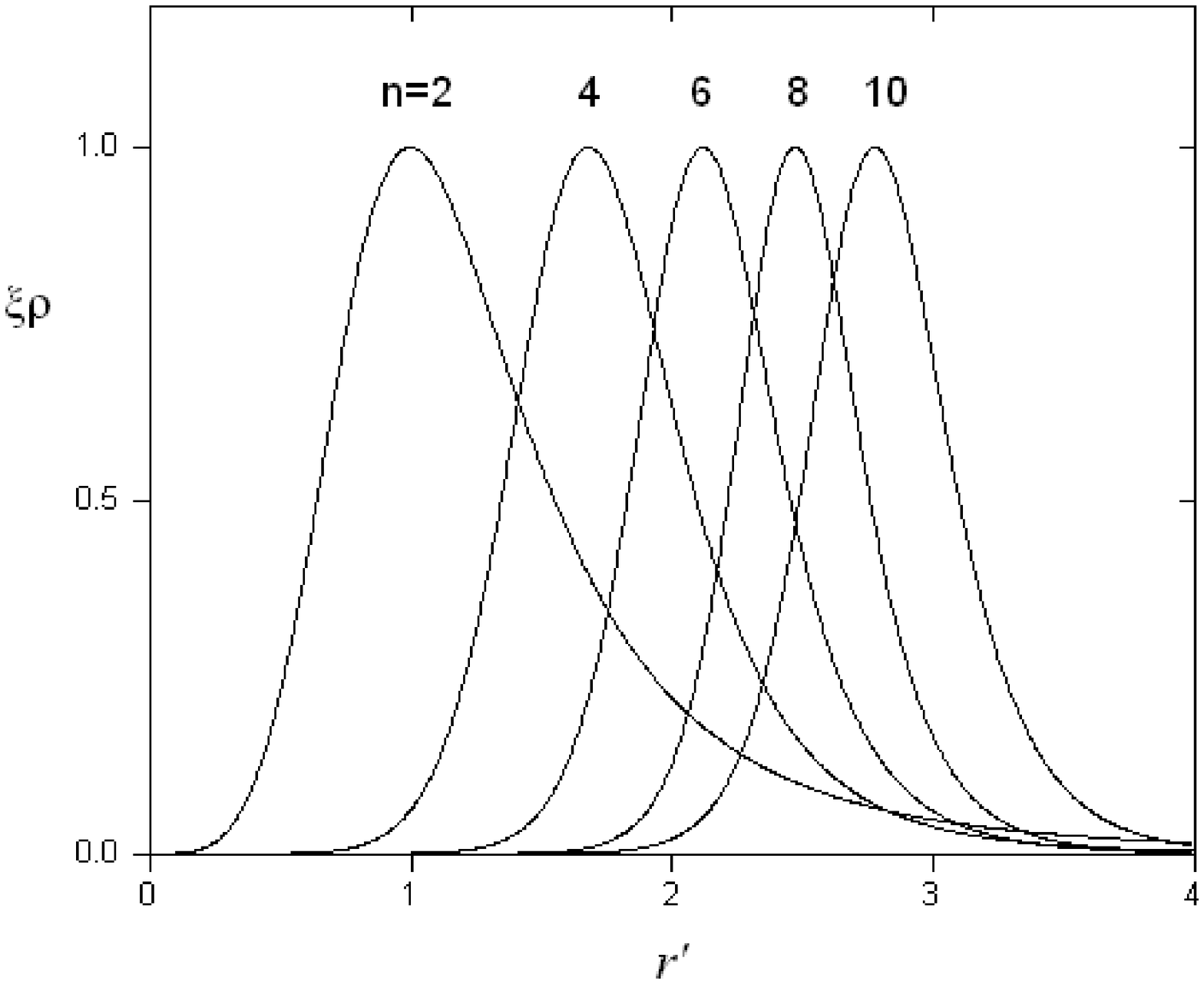} }}\rotfinish}}
{\vspace*{-8cm}} {{\hspace*{9.5cm}\rotstart{0 rotate
}{\scalebox{.35}{\includegraphics{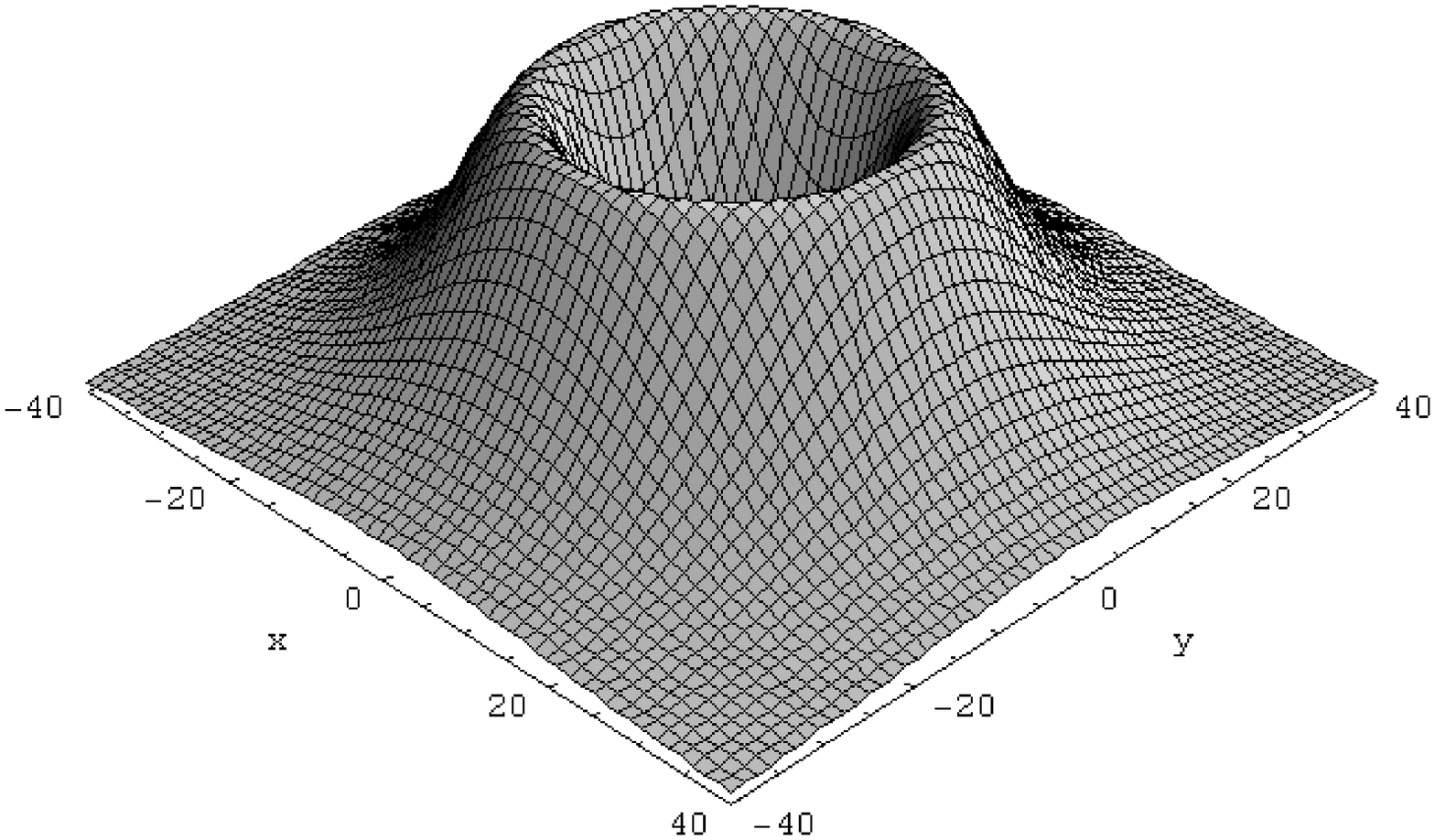} }}\rotfinish}}

\vspace{2cm}

\begin{minipage}[t]{8cm}\caption{{\footnotesize Plot of the dimensionless profile $\xi\,\rho$ as a
function of the dimensionless distance from the vortex core
$r^\prime=r/r_2$ with $r_2=\sqrt{\xi\,N}/\pi$ for the vortices
with $n=2,4,6,8,10$. }}
\end{minipage} \hspace{1cm}
\begin{minipage}[t]{8cm}\caption{{\footnotesize 3D representation of the dimensionless profile $\xi\,\rho$
for the ground vortex $n=2$ in the 2D-dimensionless space:
($x^\prime=x/r_2,\,\,y^\prime=y/r_2$) with
$r_2=\sqrt{\xi\,N}/\pi$.}}
\end{minipage}
\end{figure}
Figure 1 shows the plot of the dimensionless quantum velocity
$v^\prime=v\,m\,r_2/\hbar$ with $r_2=\sqrt{\xi\,N}/\pi$, obtained
by combining Eqs. (\ref{vq}) and (\ref{sol1}), as a function of
the dimensionless distance from the vortex core $r^\prime=r/r_2$
for the vortices with $n=2,4,6,8,10$. We observe that ${\bfm v}$
is equal to zero for $r=r_n$ where $\rho$ reaches its maximum
value. Figure 2 reports the behaviour of the non singular part of
the dimensionless vorticity
$\omega^\prime=\omega\,m\,r_2^2/\hbar$, given by (\ref{shape1})
for the same vortices of figure 1. Figure 3 shows the
dimensionless profile $\xi\,\rho$ as a function of $r^\prime$ for
the same vortices of the previous figures. Finally, a 3D
representation of the profile $\xi\,\rho$ for the ground vortex
$n=2$ in the 2D-dimensionless space:
($x^\prime=x/r_2,\,\,y^\prime=y/r_2$) is reported in Figure 4.

The energy $E=H$ of the n-vortex can be calculated easily by
substituting (\ref{sol1}) into (\ref{HJ}) and performing the
integration
\begin{eqnarray}
E=|n|\,\frac{\hbar^2 k^2}{2\,m} \ \ \ \ ;\ \ \ \
k^2=\frac{8\,\pi}{3\,\xi} \ \ \  .\label{en1}
\end{eqnarray}
The energy of the system results to be quantized and the energy
spectrum lower bounded.

We recall that the family of the vortices (\ref{sol1}) is related
to the vortices $\rho_{_{JP}}(r)$ of ref. \cite{Jackiw} through
$\rho(r)\propto r^2 \rho_{_{JP}}(r)$. We remark that the vortices
$\rho_{_{JP}}(r)$ are obtained as self-dual static solutions of a
Chern-Simons model and correspond to the same energy state with
energy equal to zero. On the contrary, for the vortex family
(\ref{sol1}), we have that any vortex corresponds to a different
state of the system whose energy is given by Eq. (\ref{en1}).

We conclude by noting that very recently \cite{KSPB}, it has been
considered a generalization of the present model (describing
neutral particles which obey the EIP) in the cases of non
relativistic charged particles. In this modified model the matter
field obeying the EIP is minimally coupled to a gauge field whose
dynamics are described within the frame of the Chern-Simons
picture. The model is a canonical one and admits self-dual static
non topological vortex solutions with zero energy and linear
momentum. The expressions of the main physical quantities
associated to these solutions are obtained. The electric charge
and the angular momentum are derived analytically, while the
shape, together with the electric and magnetic fields of the
vortex, are obtained numerically. This model can be considered as
a continuous deformation of the Jackiw and Pi one \cite{Jackiw},
performed by the parameter $\kappa$ which takes into account the
EIP.

\vfill\eject

\end{document}